# The Need for Plasma Astrophysics in Understanding Life Cycles of Active Galaxies
*A White Paper submitted to Galaxies across Cosmic Time (GCT)*

H. Li (LANL), J. Arons (UCB), P. Bellan (Caltech), S. Colgate (LANL), C. Forest (UW-Madison), K. Fowler (UCB), J. Goodman (Princeton), T. Intrator (LANL), P. Kronberg (LANL), M. Lyutikov (Purdue), E. Zweibel (UW-Madison)

**Abstract:** In this White Paper, we emphasize the need for and the important role of plasma astrophysics in the studies of formation, evolution of, and feedback by Active Galaxies. We make three specific recommendations: 1) We need to significantly **increase the resolution of VLA**, perhaps by building an EVLA-II at a modest cost. This will provide the angular resolution to study jets at kpc scales, where, for example, detailed Faraday rotation diagnosis can be done at 1GHz transverse to jets; 2) We need to build **coordinated programs among NSF, NASA, and DOE to support laboratory plasma experiments (including liquid metal)** that are designed to study key astrophysical processes, such as magneto-rotational instability (origin of angular momentum transport), dynamo (origin of magnetic fields), jet launching and stability. Experiments allowing access to relativistic plasma regime (perhaps by intense lasers and magnetic fields) will be very helpful for understanding the stability and dissipation physics of jets from Supermassive Black Holes; 3) Again through the coordinated support among the three Agencies, we need to invest in **developing comprehensive theory and advanced simulation tools to study the accretion disks and jets in relativistic plasma physics regime**, especially in connecting large scale fluid scale phenomena with relativistic kinetic dissipation physics through which multi-wavelength radiation is produced.

## I. Introduction

While the last Decadal Review hailed the discovery of supermassive black holes (SMBHs) as one of the major advances in astronomy at the turn of the $21^{st}$ century, the understanding of these SMBHs, their formation, evolution, and feedback, continues to be a frontier challenge in Astronomy and Astrophysics. In addition to the prodigious amount of radiation and energetic winds from the nuclear region, a significant fraction of the gravitational energy released during the SMBH formation can come out as enigmatic yet powerful jets and radio lobes (e.g., Burbidge 1958; Kronberg et al. 2001). These jets and lobes serve as arguably one of the most important non-thermal energy sources in the overall cosmic energy flow (e.g., Furlanetto & Loeb 2001). The active galaxies are believed to be important in shaping the galaxy formation and populations (e.g., cosmic "downsizing" effect, e.g., Shankar et al. 2009), affecting the properties of the intra-cluster medium (ICM) and the inter-galactic medium (IGM) (e.g., McNamara & Nulsen 2007), and contributing to the extra-galactic magnetic fields and cosmic rays (CRs), including ultra-high energy cosmic rays (UHECRs; e.g., Abraham et al. 2008).

In this White Paper, we wish to emphasize the essential role of Plasma Astrophysics in the study of Life Cycles of Active Galaxies. We present the need for a concerted three-pronged approach, namely building advanced observational capabilities, conducting designated laboratory plasma physics experiments that focus on understanding key astrophysical processes, and developing fundamental theory and simulations of relativistic plasma astrophysics that connect fluid physics with microphysical dissipation physics. For example, the current study of jets and lobes is mostly limited to magnetohydrodynamic (MHD) fluid description. While successful, it unfortunately neglects (often erroneously) microscopic kinetic details of particles and waves that are responsible for acceleration of particles, including possibly the UHECRs. Our lack of understanding of the underlying microphysics is a major impediment to our progress in this branch of astrophysics, especially in making consequential connections with observations.

Some of the major questions relevant to Life Cycles of Active Galaxies include: *How do SMBHs form, especially at high redshifts (z~6)? What controls the angular momentum transport processes in AGN disks? Is there an accretion disk dynamo and how does it work? How are jets accelerated and collimated? What is the relative role between BH spin and accretion disk with coronae in launching the jets? What is the jet composition? How are particles accelerated in the jet? How is the high-energy radiation generated? What determines the jet stability over kpc to Mpc scales? How do radio lobes form and what are their physical properties? Could UHECRs come from the BH-disk-jet-lobe systems? How do jets interact with the ICM and IGM? How are magnetic fields and CRs in radio lobes dispersed in the ICM and IGM? Are they the main source of magnetic fields in the ICM and IGM? Is galaxy formation predominantly an MHD process out of a magnetized IGM?* Because the mean free path of particles in these systems is often comparable to the entire system scale and magnetic fields are believed to be prevalent, these important astrophysics questions cannot be fully addressed without understanding key plasma physics processes. There has been steady progress in observations that are providing more and better constraints on the physical parameters in these systems. Through a confluence of advances in astronomical observations, fundamental theories, large-scale simulations, and laboratory experiments, we may be collectively on the verge of making significant progress in understanding this inter-related BH-disk-jet-lobe-ICM-IGM-galaxy formation sequence.

## II. Plasma Astrophysics in Studying Active Galaxies

Some of the specific areas where we believe substantial progress can be made in the coming decade and where plasma astrophysics plays a pivotal role include:

**A) Formation of SMBHs: What Role do Magnetic Fields Play?**

The process of angular momentum transport (AMT) and the critical role of magnetic fields in accretion disks share a considerable similarity with accretion onto stellar mass black holes and neutron stars. We refer to a White Paper submitted to the SSE panel from GPAP and CMSO for a thorough discussion. One prominent unresolved issue in SMBH systems is the possibility of an efficient accretion disk dynamo. In particular, the synergy among theory, computer simulations, and the laboratory dynamo and AMT experiments (e.g., Spence 2006; Ji et al. 2006) makes it a very exciting possibility that a substantial progress can be made in the coming decade.

## B) Relativistic Jets from SMBHs: Theory and Simulations of Relativistic Extended MHD and Relativistic Kinetic Plasmas

The jet's composition, formation, collimation, stability, and relativistic particle acceleration mechanisms remain as some of the most outstanding, unsolved issues. In the past, jets have been studied in the ideal fluid limit (both hydromagnetic and Poynting flux limits; e.g., Blandford and Payne 1982; Blandford 1976; Lovelace 1976; Lynden-Bell 1996; Li et al. 2006). One major development during the past decade is the General Relativistic ideal MHD simulations of the dynamics near SMBHs and the production of relativistic outflows (e.g., McKinney & Narayan 2007). Several possible directions of research are:

• **Is the jet launching an essentially electrostatic process or an hydromagnetic process?** Mass loading in launching jets is critical in determining jet's composition, acceleration, and collimation. This requires a detailed understanding how the plasma transitions from the collisional state in the disk, to a (possible) tenuous, collisionless coronae, to a relativistic collisionless outflow. The large electric potential drop that exists between SMBH and disk (and between different disk radii) needs to be properly taken into account, most likely not in the ideal MHD limit. We need relativistic kinetic theory to understand how plasmas organize themselves to respond to the BH-disk driving (including the driving by the spin of SMBHs).

• **Properties of relativistic flows in the strong field limit:** Plasmas in the vicinity of SMBHs may differ from the more conventional laboratory and near-space plasma. First, plasma inertia may become dominated by magnetic fields and not by the particle rest masses $\rho$, so that the $B^2/(8\pi c^2) \geq \rho$. Second, when plasma motion is relativistic, the displacement current $(c/4\pi)\partial_t E$ then becomes of the order of the conduction current $j$ (in a non-relativistic plasma displacement current is rarely important). Similarly, charge density $\rho_e$ may become of the order of $j/c$. (Similar issues arise in pulsar and magnetar magnetospheres.) Some of the basic theoretical understanding on how such plasmas behave still needs to be worked out.

• **What determines the global jet/lobe structure, especially in the relativistic regime?** Though it is widely believed that the jets are produced in the Poynting dominated limit, it is not clear whether an electro-magnetic description (with a strong axial current, for example) will persist to the global scale. This has important implications for understanding the AGN feedback in galaxy clusters, for example. Do jets necessarily become kinetic energy dominated flow beyond tens of kpc? What regulates this conversion process?

• **Why are jets quasi-stable and collimated up to Mpc?** More effort needs to be devoted to understanding jet stability in the relativistic limit. Furthermore, we need to seriously ask the question whether ideal MHD is enough to describe jet flow or whether we must consider including extended MHD effects, such as heat flow parallel to the background magnetic fields and pressure anisotropy. Most previous relativistic MHD research is still based on the use of a stress tensor whose origin is based on thermodynamic considerations rather than electrodynamics, in which electromagnetic forces dominate. The latter might be closer to reality in the BH-disk-jet-lobe system. Casting the jet/lobe system in an extended relativistic MHD framework may prove fruitful and even necessary in addressing the jet stability problem.

• **How are particles accelerated in jets and how do they produce the observed radiation?**
The coming decade offers a great opportunity in particle acceleration studies, given the anticipated observations from Auger, Fermi GST, and EVLA, among others. The traditional four mechanisms for particle acceleration, namely, $2^{nd}$ order Fermi, shocks, electostatic (double) layers, and reconnection, all need to be extended to relativistic regimes. The understanding of reconnection in the relativistic environment has hardly begun. In systems where the magnetization is not as large, the dissipation, particle acceleration and resulting photon emission are usually ascribed to acceleration at relativistic shock waves - these are inefficient in high σ ($=B^2/4\pi\gamma\rho c^2$) flows. Substantial development has gone into migrating the Diffusive Shock Acceleration mechanism to the relativistic environment. Overall, this kind of understanding is crucial for connecting the observations to the system models in a quantitative manner. Further advances in computational technique in the coming decade will enable substantial computational efforts in these challenging areas.

**C) Fate of Radio Lobes and Feedback to the ICM and IGM:**

The fate of the giant radio lobes has become an increasingly important question both in understanding the origin of magnetic fields in the ICM and IGM, and in assessing their dynamic influence on cosmic structure formation. Furthermore, the multi-wavelength observations of radio jets/lobes inside X-ray clusters have provided unprecedented information on how the jets/lobes interact with the surrounding plasmas (e.g., McNamara & Nulsen 2007), which in turn provides valuable constraints on the nature of jet/lobes (e.g., Diehl et al. 2008). In addition, numerous laboratory experiments have been conducted or can be used to study some of the relevant physical processes in the lobe-ICM interactions (e.g., Liu et al. 2008). These advances are leading us into the next stage of jet-lobe physics understanding and modeling. Several possibly fruitful areas of research include:

• **Will magnetic self-organization play a determining role in controlling the lobe expansion?**
With most of their magnetic energy still intact in $\sim 10^8$ yrs, the long term evolution of the lobes in $\sim 10^{10}$ yrs has important impact and implications for magnetizing the ICM and the IGM. What is the expansion rate of the lobes? What is the dynamics of self-organization during expansion? Will the magnetic flux surfaces be destroyed due to the interaction (e.g., reconnection) with the pre-existing (weak) background magnetic fields? How does the plasma inside the lobe leak out of the lobe? Could the lobes be the primary source of the extra-galactic cosmic rays from GeVs to PeVs and above? Comprehensive 3-D MHD simulations, being guided by observations and validated by laboratory plasma experiments could help address these questions.

• **Dynamics of asymmetric magnetic reconnection between the lobe and the background medium.** The physical condition inside and outside the lobes can be very different (relativistic vs. non-relativistic, possibly pair plasmas vs. e-p plasmas, etc.). Magnetic reconnection in this regime has not been carefully studied, though asymmetric reconnection in the non-relativistic plasmas has been observed both in the magnetospheric plasma via THEMIS satellite (e.g., Mozer et al. 2008) and the laboratory experiments such as between two kinking flux ropes (e.g., Intrator et al. 2009). Recent advances in magnetic reconnection theory and simulations could be applicable, though additional theoretical and computational development is necessary. For example, how would turbulent cascade develop in relativistic plasma? Do current sheets have similar properties in relativistic plasmas as in non-relativistic plasmas? What controls the reconnection rate in relativistic plasmas? Advanced Particle-in-Cell simulations are poised to make headways in these areas. Ultimately, such processes will determine the lifetime and effectiveness of radio lobes in sharing their energy with the surroundings.

**D) Acceleration of UHECRs**

The recent results by the Pierre Auger Observatory on the confirmation of the so-called GZK cut-off and the possible association of UHECRs with nearby AGNs have given strong support to the idea that the SMBH-disk-jet-lobe system could be responsible for UHECRs (e.g., Abraham et al. 2008; Lyutikov & Ouyed 2007). Once again, we have to move beyond the ideal fluid approach in order to describe the overall jet-lobe dynamics together with the acceleration of energetic particles in a self-consistent manner. Currently, there is no known relativistic kinetic MHD theory and formulation that can be readily applied to model the fluid jet with energetic CRs in the jet. This points to the need to develop innovative theoretical and numerical techniques that can link relativistic fluid modeling with the relativistic kinetic (such as PIC) together. Astrophysicists, helped by laboratory plasma physicists and space plasma physicists, must take up this challenge.

## III. Ties with Observations and Laboratory Experiments

Extensive observations of AGN jets/lobes (morphology, spectra, power, field configuration, velocity, environments, etc.), as well as the X-ray cavities in galaxy clusters (size distribution, interaction properties, heating and dissipation, etc.), will provide crucial guidance and constraints on all the modeling efforts that try to incorporate both fluid and kinetic effects described above. Close collaboration with laboratory experimental efforts is necessary to validate the basic theoretical framework and numerical simulations in relevant regimes. During the past decade, there have been a small but increasing number of dedicated plasma laboratory experiments for investigating astrophysical processes. This trend must be continued and supported.

**A) Observations: A Case for EVLA-II**

Opportunities exist for exploiting X-ray, optical, and radio imaging from 1" to sub-milliarcsec with instruments that will be available in the next decade. Quite significant for jet studies is a currently shelved enhancement plan for the currently upgraded EVLA. Perhaps the single most important impact could come from significantly improving the angular resolution of kpc-scale jets. Specifically, a factor of 10 or more resolution is needed *transverse* to the jets. This improved resolution, to ~ 100 milli-arcsec, is especially important near 1 GHz, where detailed Faraday rotation diagnosis can be done, giving local magnetic field geometries in 3-D, and possibly magnetic configuration variations over a few years. Adding a few more radio telescopes within New Mexico would give enormous incremental bang for buck (over the EVLA) for an extra < $150M it would cost. This enhancement, originally proposed as the ``EVLAII'', would also serve other Science areas.

**B) Laboratory Experiments**

While many experiments are exploring the non-relativistic plasma regime, such as the on-going dynamo, MRI, spheromak-jet (e.g., Bellan et al. 2005; You et al. 2005), reversed field pinch-flux conversion, there are several laboratory regimes where relativistic behavior is important. We need to leverage this experience with relativistic plasmas to astrophysical situations.

• **Relativistic plasmas in the laboratory:** Examples of laboratory plasmas where relativistic behavior was dominant include a toroidal confinement device where a relativistic electron ring provided the entire toroidal current and was so strongly diamagnetic as to reverse the initial magnetic field (e.g., Fleischmann et al. 1976), Tokamak runaway electrons that are accelerated to MeV energies by going many times around a loop having only a 0.1 volt drop (e.g., Panaccione et al. 2003), and an electron cyclotron resonant mirror device where microwave-heated electrons at relativistic energies produced a strong diamagnetic dip in the background magnetic field (e.g., Hosokawa et al. 1993). Furthermore, it is possible to use intense lasers to generate electron-positron pairs with relativistic temperature. The upcoming NIF lasers could produce outflows that are relativistic for electrons. In addition, it is possible to design experiments to explore the regime that magnetic energy density is relativistically dominant (though perhaps without flows). This will provide valuable experimental data where relativistic MHD and kinetic theories and simulations can be tested and validated.

• **Collisionless MRI and dynamo experiments:** One further development in the MRI and dynamo areas is to consider effects beyond MHD that may occur in collisionless plasmas, that may be very different than that of the well-studied single fluid MHD (e.g., Quataert et al. 2002). In collisionless plasmas, like hot accretion disks, bending of magnetic field lines can cause the plasma pressure to become increasingly anisotropic. This feedback can modify the linear stability of the MRI. New experiments investigating plasma flow driven instabilities may be the only way to begin to address the microphysics of the MRI, and several experiments are beginning in this area (e.g., Wang et al. 2008). In particular, comparisons between experiments that use liquid metal and those that use plasma may elucidate these uniquely plasma effects.

Specifically for dynamo experiments, it is still important to experimentally explore less turbulent dynamos for a possible demonstration of astrophysical dynamos that might explain the large scale magnetic fields of planets, stars and AGN. Furthermore, there is a possibility of studying dynamos that builds on excitement in recent years of liquid metal dynamo experiments (Spence et al. 2006; Nornberg et al. 2006): un-magnetized liquid metals have been mechanically stirred and magnetic fields spontaneously created and observed. A plasma experiment has the potential to extend these studies to more astrophysically relevant parameters. Beyond the obvious fact, that most naturally occurring dynamos are plasmas, the use of plasma, rather than liquid metals to study magnetic field generation will allow the magnetic Reynolds number to be more than a factor of 10 larger than in liquid metal experiments. It will also allow the viscosity to be varied independently of the conductivity: the magnetic Prandtl number (the ratio of the magnetic Reynolds number to the hydrodynamic Reynolds number) can be varied from the liquid metal regime ($<< 1$) to the regime ($>>1$) thought to be a critical parameter that governs the nature of many astrophysical situations since it governs the onset and nature of the turbulence. Finally, advanced plasma diagnostics will be applied, for the first time, to dynamo studies. Such laboratory plasmas have never before been studied (most hot, fast flowing plasma experiments have been magnetized), but recent advances in permanent magnet technology, plasma source development and new scheme for driving flow now make it possible (Spence et al. 2009).

## III. Conclusion

The study of Active Galaxies has increasingly called for more advanced plasma astrophysics, some of which are even beyond the current state of knowledge in plasma physics, such as the need to develop relativistic fluid theory while including relativistic kinetic physics in a self-consistent fashion. On the other hand, during the past decade, it is very encouraging to see the rigorous development of laboratory plasma experiments that are, in many cases, specifically designed to study key astrophysical processes, such as dynamo, MRI, magnetic reconnection, jet dynamics, turbulence, relaxation of lobes, etc. These experiments not only provide direct physical understanding of key astrophysical processes, but also furnish valuable data for validating astrophysical simulation tools that are used to explore regimes beyond what terrestrial experiments can do.

The confluence of laboratory experiments, astronomical observations, basic plasma astrophysics theory, and sophisticated computer simulations will usher us into a new era of studying Active Galaxies with new tools and approaches provided by plasma astrophysics.